\documentclass[sigconf, nonacm]{acmart}

\usepackage{subcaption}
\usepackage{footmisc}
\usepackage{makecell}

\AtBeginDocument{
  \providecommand\BibTeX{{
    \normalfont B\kern-0.5em{\scshape i\kern-0.25em b}\kern-0.8em\TeX}}}

\begin{document}

\title{Fact Checking Chatbot: A Misinformation Intervention for Instant Messaging Apps and an Analysis of Trust in the Fact Checkers}

\author{Gionnieve Lim}
\email{gionnievelim@gmail.com}
\orcid{0000-0002-8399-1633}
\affiliation{
  \institution{Singapore University of Technology and Design}
  \streetaddress{8 Somapah Rd}
  \country{Singapore}
  \postcode{487372}
}

\author{Simon T. Perrault}
\email{perrault.simon@gmail.com}
\orcid{0000-0002-3105-9350}
\affiliation{
  \institution{Singapore University of Technology and Design}
  \streetaddress{8 Somapah Rd}
  \country{Singapore}
  \postcode{487372}
}

\renewcommand{\shortauthors}{Lim and Perrault}

\begin{abstract}
In Singapore, there has been a rise in misinformation on mobile instant messaging services (MIMS). MIMS support both small peer-to-peer networks and large groups. Misinformation in the former may spread due to recipients’ trust in the sender while in the latter, misinformation can directly reach a wide audience. The encryption of MIMS makes it difficult to address misinformation directly. As such, chatbots have become an alternative solution where users can disclose their chat content directly to fact checking services. To understand how effective fact checking chatbots are as an intervention and how trust in three different fact checkers (i.e., Government, News Outlets, and Artificial Intelligence) may affect this trust, we conducted a within-subjects experiment with 527 Singapore residents. We found mixed results for the fact checkers but support for the chatbot intervention overall. We also found a striking contradiction between participants’ trust in the fact checkers and their behaviour towards them. Specifically, those who reported a high level of trust in the government performed worse and tended to follow the fact checking tool less when it was endorsed by the government.
\end{abstract}

\keywords{Misinformation, Automated fact checking, Trust, Instant messaging, Chatbot}

\maketitle

\section{Introduction}

Defined as the “inadvertent sharing of false information” \citep{Wardle2017}, misinformation has attracted much attention following the 2016 US presidential election \citep{KF2018} when “fake news” became a popular term used by media around the world. Misinformation is spread through information and communication technologies such as blogs, forums, mobile instant messaging services (MIMS), and social media platforms. Given the ease of access and communication made possible by highly interconnected services that allow nearly instantaneous creation and exchange of information, misinformation has become a common phenomenon \citep{Meel2020, Zhou2020}. When used by malicious parties for hostile purposes, misinformation becomes a tool to drive extreme polarisation in democracies, stirring crime and conflict \citep{Carothers2019}. Misinformation can cause harm in other ways, such as stirring confusion, fear, unease and panic among people as in the case of the COVID-19 infodemic \citep{WHO}.

As a digitally connected country with a diverse population, Singapore is not immune to the problem of misinformation \citep{Lim2019}. There have been multiple instances of misinformation over the years. In 2015, a Singaporean teen fabricated a Prime Minister’s Office webpage that announced the passing of the country’s first prime minister, Mr. Lee Kwan Yew, to show how easy it was for a hoax to spread. This led to false reporting of Mr. Lee’s death by foreign news media which later retracted the news \citep{Philomin2015}. In 2017, a police raid conducted in a bazaar was falsely attributed to the sale of non-Halal food\footnote{Non-Halal food do not follow the dietary observances of Islamic law and Muslims are prohibited from consuming them.}. An official address was posted by a Member of Parliament on Facebook to allay fears, particularly within the Muslim community, clarifying that the raid was targeted at unlicensed foreign food handlers working illegally at the bazaar \citep{Today2017}. Since early 2020, the infodemic accompanying the COVID-19 pandemic also took root in Singapore. There were rumours on the locations of the infections, instances of deaths caused by the virus, measures imposed by the government \citep{GovSG2020} and the side effects of COVID-19 vaccines \citep{MOH}.

A significant volume of misinformation can be found in MIMS that are used for daily and largely personal communication \citep{Pasquetto2020, Ng2021, Nobre2022}. As of 2020, WhatsApp is the most popular MIMS in Singapore, used by 87.1 per cent of internet users, followed by Facebook Messenger (53.2 per cent), WeChat (32.5 per cent) and Telegram (30.1 per cent) \citep{Kemp2021}. Due to the high adoption of MIMS that expose the population to the threat of misinformation, countermeasures that can be implemented on the platforms become more pertinent. In this study, we conducted an experiment with 527 Singapore residents to understand the effectiveness of a MIMS chatbot intervention that provided a fact checking service. We sought to understand how effective the chatbot is in affecting people’s perceptions of the veracity of news and how trust in three different fact checkers (i.e., Government, News Outlets, and Artificial Intelligence) affect that trust.

\section{Related Work}

\subsection{Misinformation on MIMS}

Cheap smartphones and data plans have made MIMS accessible to many. With added functionalities that support the formation of group chats and use of multimedia formats, MIMS have greatly enhanced communication, both in volume and in variety. However, the ease of access and connectivity, coupled with their closed and encrypted nature, have created an environment suitable for misinformation to foster \citep{Pasquetto2020, Gursky2021}. With the forwarding function, misinformation is amplified when a bogus message is sent from one person to another, particularly in group chats \citep{Tandoc2020}. The fact that content can be easily propagated without any association with the original context or sender further reduces accountability \citep{Pellegrino2018}. The propagation of misinformation can lead to tragic consequences. An example is a series of unrelated lynching events that occurred in India due to rumours of child kidnappers that were shared on WhatsApp \citep{BBC2018}.

In Singapore, misinformation has been detected in various MIMS. In the tracing of COVID-19 misinformation in a Telegram group chat that had over 10,000 participants, 72 pieces of misinformation were found in which many had not been publicly addressed \citep{Ng2021}. On WhatsApp, misinformation covers a wide variety of topics including government policies \citep{Ang2019, Iau2019}, crime and safety \citep{Menon2019, Yong2019}, and COVID-19 \citep{Goh2020}. In 2018, WhatsApp was found to be the most common source of fake news encountered by Singapore residents \citep{Sin2018} and was reported as a media for news consumption by 44 per cent of Singaporeans \citep{Ipsos2018}. While a majority of Singapore residents are confident in their ability to spot fake news, nearly half also admit to having fallen for fake news before \citep{Ipsos2018}. The disparity between the perceived and actual ability to discern the veracity of news was also established in a study which found that more than twothirds of the Singaporean respondents failed to correctly identify that a manipulated news article was untrustworthy \citep{IPS2020}. The findings suggested that information literacy among Singaporeans was low as many respondents were unable to recognise the many signs of manipulation in the article.

\subsection{Interventions on MIMS}

Several initiatives have been introduced to counter misinformation on MIMS. WhatsApp and Facebook Messenger have limited message forwarding in a bid to curb extensive spreading of misinformation \citep{WhatsApp}, although its effectiveness remains questionable. A study found that while the intervention can delay spread, it is not effective in preventing viral content from quickly reaching an extensive network \citep{Melo2020}. WhatsApp also introduced “frequently forwarded” labels for messages that have been sent more than five times to indicate that those messages have been disseminated by many \citep{WhatsApp}. WeChat has a dedicated section for fact checked information \citep{Shen2020}. Beyond commercial initiatives, academics have also explored a range of strategies to mitigate misinformation. They include using machine learning to automate the fact checking process \citep{Hassan2017}, adding credibility indicators to posts to inform users on the veracity of the contents \citep{Gao2018}, and using nudges, such as in the form of reminders, to encourage mindful sharing \citep{Pennycook2020}.

While numerous measures have been implemented on MIMS, most of them do not directly address the content of the misinformation and fall short on dealing with the misinformation with immediacy and certainty. For instance, limits on message forwarding are ineffective \citep{Melo2020}, forward labels are merely suggestive of intentional dissemination and a dedicated fact checked section may not contain the latest misinformation. The use of these roundabout measures can be explained by MIMS serving mainly as a private communication tool for which end-toend encryption protects the privacy of the content exchanged among users \citep{WhatsApp2, Telegram, LINE2015, Kricheli2021}. Unlike social media where much content is shared in the public domain, the content posted on MIMS is not readable by a third party. Alternative solutions to work around this restriction have emerged. One technique is the tracking of metadata such as the unique hash of content that has been flagged by the community to prevent further dissemination of it \citep{Gupta2018}. Fact checking organisations have also leveraged the chatbot functionality in MIMS to provide fact checking services \citep{Grau2020, Mantas2021}. Users can send the message they are suspicious of through a chatbot, thereby directly disclosing the private content to a third party for fact checking. This facilitates quicker evaluation and warning of misinformation. As automated fact checking technologies develop, the chatbot solution can be an effective tool to counter new misinformation that has not been assessed in time by professional human fact checkers.

There has been a particular rise in the usage of chatbots to address the infodemic surrounding COVID-19. Studies have assessed the use of chatbots to answer queries \citep{El2021, Gunson2021, Roque2021}, detect misinformation \citep{Della2018} and debunk falsehoods \citep{Pecher2020, Yu2018}. The evaluation of some of these chatbots are generally positive. In a comparison study on a set of news chatbots managed by various international news organisations, news chatbots that provided relevant, diverse and up-to-date information and responded with immediacy and with human traits were preferred by participants. In a study conducted in Saudi Arabia, despite a majority of participants being unaware of health chatbots, they had positive perceptions towards the chatbot used in the study as they found it functional and useful \citep{Almalki2020}. A study on a question-answering chatbot (“BotCovid”) found satisfaction among users who had positive perceptions towards its functionality, compatibility and reliability \citep{Roque2021}. A study on a healthcare chatbot (“Chasey”) reported that participants perceived it to be non-complex, useful and satisfying and they would recommend the chatbot to others \citep{El2021}. While we did not find studies focusing solely on fact checking chatbots, the aforementioned studies point to the general acceptance and usability potential of the chatbot as an information delivery system for crucial content that are of public interest. For instance, a chatbot by the International Fact Checking Network at Poynter Institute (“FactChat”), sent 500,000 messages that served 82,000 people in the months preceding the 2020 US presidential election \citep{Mantas2020}, demonstrating the feasibility of the chatbot as a misinformation intervention.

\subsection{Trust in Fact Checkers}

Different media sources and news formats are met with different levels of trust by people. Media trust is often associated with media credibility \citep{Kiousis2001}. Trusting involves a degree of risk and uncertainty and people rely on credibility clues to validate their choice of trusted media sources \citep{Kohring2019}. These credibility clues include expertise, trustworthiness, fairness, bias and accuracy of the source, among others \citep{Hovland1953, Gaziano1986, Meyer1988, West1994}. With trust, people engage more and become less sceptical of news content from the media sources when seeking information \citep{Fletcher2017}, making the reliability of the sources an important consideration.

In the context of fact checking, trust plays a similar role. The purpose of fact checking is to give credence to facts and to debunk falsehoods. As they serve to inform, fact checks bear similar characteristics to news, the difference being that they are secondary reports. Where trust is concerned, the fact checker providing the service is of key consideration. If the fact checker is not trusted, the fact check becomes pointless as no amount of evidence will be deemed reliable \citep{Brandtzaeg2017}. This puts to waste the resources used to collate and organise evidence as part of the fact checking process. Furthermore, people may turn to alternative channels that are of low credibility, becoming more vulnerable to misinformation \citep{Ladd2012}. Fact checking services in Singapore are provided by several fact checkers—the government, news outlets and fact checking groups \citep{NLB}. There have also been investments made by the government to automate the process of misinformation identification \citep{AISG2021}.

There have been studies that examined people’s perceptions and attitudes towards fact checkers, albeit with an overwhelming focus on Western democracies where societies see more polarising perspectives. For instance, government entities are perceived as authoritative and reputable by those who trust them, while those who do not are sceptical about the information the government entities present or may possibly withhold. News outlets are also thought to either convey knowledge or create sensationalism \citep{Glasdam2020}. Fact checking organisations are seen as reliable and useful by those who trust them and thought of as lacking expertise and integrity by those who do not \citep{Brandtzaeg2018}. In Singapore, trust levels in the government and news outlets are generally high. A study found that the most trusted sources by Singapore residents were television, print newspapers of the mass media and radio, while the least trusted sources were online discussion forums, MIMS and social networking sites \citep{IPS2020}. In another study, government communication platforms were reported as the most trusted source, among 11 information sources \citep{Lim2021}. Trust in the source plays an important role in the acceptance and, subsequently, impact of the information given by the source \citep{Pornpitakpan2004, Walter2021}. A fact checker that is trusted will be more greatly relied upon to provide a clean information space and understanding the differences can signal which services demand more attention and resources.

However, some adverse effects of trust, pertaining to blind trust and over trust, have also been observed and they call for caution when assimilating information given by fact checkers. For instance, a study observed that greater trust in politicians also empowers them to lie and avoid being held to public scrutiny for their statements and actions on issues that they are perceived to be more competent and capable of addressing \citep{Ceron2021}. In Singapore, where there is high trust in the government, this could be a potentially great pitfall if there are failures of integrity in state practices and communications. Also, in a study on an automated fact checker, it was observed that participants were often misled to follow wrong predictions given by the system, suggesting that they were overly trusting of it \citep{Nguyen2018}. When there is too much trust in the source, the accuracy of the content may be implicitly taken as true and hence overlooked. While one would expect a trusted source to deliver reliable information, this also becomes a drawback when taken advantage of. This highlights the importance of being sceptical and critical \citep{Vraga2021} of information that comes even from a trusted source.

\section{Method}

To investigate the effects of fact check labels on news verified through an instant messaging chatbot and the trust in the fact checkers, we conducted a within-subjects experiment and a post-experiment survey. The experiment was used to assess the effectiveness of the chatbot intervention and the different fact checkers while the survey examined trust perceptions towards the fact checkers. In the experiment, participants had to rate the authenticity of 16 news headlines (i.e., whether they were true or false) that had been fact checked and labelled (as either true or false) by various fact checkers.

\subsection{Inquiry}

In the study, we sought to answer the research question (RQ):

\begin{itemize}
    \item \textbf{RQ:} How does the level of trust in the fact checker affect the effectiveness of the fact check labels?
\end{itemize}

We posited that knowing who provided the fact checking service would influence users’ decision on whether a news item is true or false. Our hypotheses (Hs) were:

\begin{itemize}
    \item \textbf{H1:} Different fact checkers will lead to different levels of accuracy in judging the veracity of news.
    \item \textbf{H2:} Different fact checkers will lead to different levels of adherence to the fact check labels when judging the veracity of news.
\end{itemize}

\subsection{Participants}

We engaged a survey company\footnote{TGM Research (\url{https://tgmresearch.com/}) was engaged for the recruitment of participants.} to recruit participants who were 18 years old and above, fluent in English and residing in Singapore. We received 568 responses and removed duplicates\footnote{Multiple responses made by the same participant (identified by their participant identification number) were removed.} ($n = 11$) and responses with straightlining\footnote{Responses in which the participant gave identical answers to each series of questions in the experiment were removed.} ($n = 30$). In all, the study had 527 participants. See Table~\ref{tab:age} for their gender and age distributions. On the level of education, 0.2 per cent of the participants had no formal education, 4.6 per cent had primary education (Primary School Leaving Examination), 26.0 per cent had secondary education (General Certificate of Education Ordinary, Normal or Advanced Level), 26.9 per cent had vocational education (Diploma or Nitec), and 42.3 per cent had tertiary education (Bachelor, Master or Doctoral). On citizenship, 85.8 per cent are Singapore citizens, 10.1 per cent are permanent residents and 4.2 per cent are pass holders.

\begin{table}[h]
    \centering
    \caption{The distribution of the participants compared to the 2021 national population of Singapore taken from SingStat \citep{SingStat}.}
    \label{tab:age}
    \begin{tabular}{c c c}
        \toprule
        Group & \makecell{National\\representation (\%)} & \makecell{Achieved sample\\($N=527$)} \\
        \midrule
        Female & 51.0 & 52.0\% (274) \\
        Male & 49.0 & 48.0\% (253)  \\
        \midrule
        18-24 & 9.8 & 13.5\% (71) \\
        25-34 & 17.6 & 23.0\% (121) \\
        35-44 & 17.7 & 22.2\% (117) \\
        45-54 & 17.8 & 16.3\% (86) \\
        55-64 & 17.7 & 15.9\% (84) \\
        65-99 & 19.4 & 9.1\% (48) \\
        \bottomrule
    \end{tabular}
\end{table}

\subsection{Experiment}

The experiment involved four independent variables: Fact Checker, News Veracity, Fact Check Label and Label Precision and two dependent variables: Accuracy of the Perceived Veracity and Adherence to the Fact Check Label.

\subsubsection{Independent Variables}

Fact Checker referred to the provider of the fact checking service: Government, News Outlets, Artificial Intelligence and Control. In the Control condition, no fact checker was shown.

News Veracity referred to the actual veracity of news with two levels: True and False. The news veracity of the 16 news items that were used for the experiment is indicated in Table~\ref{tab:newsitems}.

Fact Check Label referred to the label applied after the fact checking of news with two levels: True and False. An equal proportion of labels were applied to the 16 news items such that for News Veracity × Fact Check Label, there were four news items each for the True × True, False × False, True × False, and False × True combinations. The latter two were “oppositely labelled” news items meant for better assessing trust.

Label Precision referred to whether the news was given the correct fact checked label with two levels: Correctly Labelled and Incorrectly Labelled. This variable was derived from News Veracity × Fact Check Label, where the True × True and False × False combinations meant that the label matched the actual news veracity and was correct, while the True × False and False × True combinations meant that the label was incorrect.

\subsubsection{Dependent Variables}

The dependent variables were derived from the main question in the experiment that sought to capture participants’ perceived veracity of news. In the experiment, participants were shown 16 news items and, for each one, had to answer an authenticity question: “How authentic do you think the news in the chat is?” on a 4-point Likert scale [Definitely False, Somewhat False, Somewhat True, Definitely True].

The Accuracy of the Perceived Veracity modified the authenticity question based on News Veracity by examining whether there was a match between the perceived and actual veracity of news. It took on a 4-point scale [1: Inaccurate, 2: Somewhat Inaccurate, 3: Somewhat Accurate, 4: Accurate]. For example, if the actual veracity of the news was “True”, and the perceived veracity response was “Somewhat False”, the accuracy of the response would be “2: Somewhat Inaccurate”. If the perceived veracity response was “True” instead, the accuracy would be “4: Accurate”.

The Adherence to the Fact Check Label modified the authenticity question based on Fact Check Label by examining whether there was a match between the perceived veracity of news and the label given by the fact checker. It took on a 2-point scale [0: Does Not Adhere, 1: Does Adhere]. For example, if the perceived veracity response was “Somewhat False” or “Definitely False” and the label was “False”, adherence would be “1: Does Adhere”. If the label was “True” instead, adherence would be “0: Does Not Adhere”. We adopted a 2-point scale as we were more interested in the polarity of their perception.

\subsubsection{Procedure}

\begin{figure}[h]
    \centering
    \includegraphics[width=0.5\linewidth]{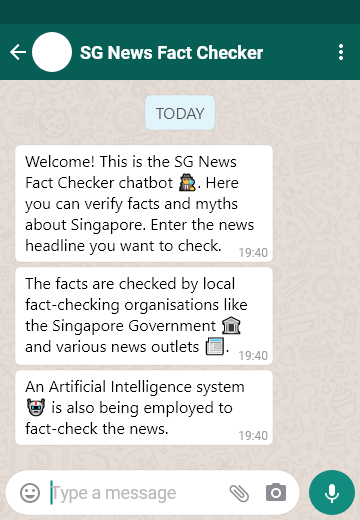}
    \caption{The introductory message describing the purpose and usage of the chatbot.}
    \label{fig:chatbot}
\end{figure}

The experiment was conducted online through a web app developed by the researchers. Participants had to pass a screening stage arranged by the survey company to receive a link to the web app. Upon gaining access, a welcome screen introduced them to the study and the chatbot interface (see Fig.~\ref{fig:chatbot}). Participants then had to answer a set of demographic questions and those who were unwilling to share their information could withdraw early on. As participants were represented by a participant identification number, no personally identifiable information was collected. The study was approved by the University’s Institutional Review Board.

During the experiment, participants had to answer an authenticity question for a randomised series of 16 news items. They were asked to rate “How authentic do you think the news in the chat is?” on a 4-point Likert scale [Definitely False, Somewhat False, Somewhat True, Definitely True]. The web app was programmed such that the Likert scale was counterbalanced across the participants where the scale would be arranged in either the [False to True] or [True to False] order for each participant randomly. Thereafter, a post-experiment survey that included several single-choice items and a ranking question was used to collect participants’ thoughts on the chatbot and the fact checkers. Upon submission, participants were redirected to a completion page provided by the survey company.

\subsubsection{Interface}

\begin{figure*}[h]
    \centering
    \includegraphics[width=\linewidth]{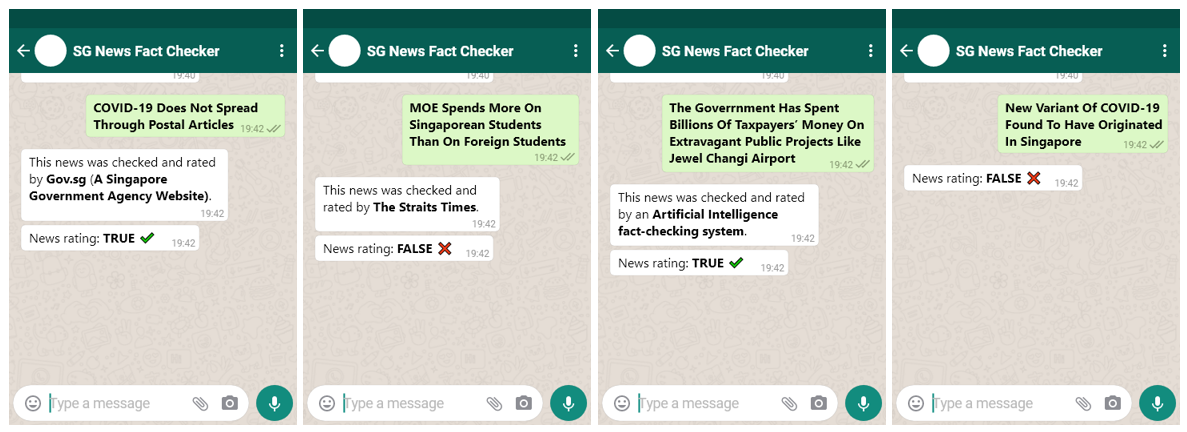}
    \caption{The mock WhatsApp chatbot interface showing conversations of four news items in different Fact Checker × News Veracity × Fact Check Label conditions.}
    \label{fig:interface}
\end{figure*}

The experiment used a screen capture of a chatbot conversation to display each news item (see Fig.~\ref{fig:interface}). For realism, the interface was designed to mimic WhatsApp. The title of the chatbot was “SG News Fact Checker”, a generic yet relevant name that was not associated with an existing account at the time of the study. A partially hidden chat bubble was added to indicate that the chatbot had a conversation history. Next is the chat bubble containing the news item where the headline was highlighted in bold to emphasise its content. Following that is the chat bubble indicating the fact checker. The final chat bubble is the fact checking result with the fact check label given by the fact checker. A “TRUE” result had a tick emoji while a “FALSE” result had a cross emoji. The Government fact checker is described as “Gov.sg (A Singapore Government Agency Website)”, News Outlets as “The Straits Times” and Artificial Intelligence as “Artificial Intelligence fact checking system”. The Control condition did not have a fact checker chat bubble. “Gov.sg” was used to represent various government departments since it is the main government communication platform \citep{GovSG}. “The Straits Times” was chosen due to its position as a leading news publisher in Singapore \citep{Yong2019b}.

\subsubsection{Stimuli}

Table~\ref{tab:newsitems} shows the 16 pieces of news headlines used for the experiment. The news items were sourced from Factually\footnote{\url{https://www.gov.sg/factually}.}, a government fact checking website. For relevance, only news that were published in the last three years at the time of the study, from 2019 to 2021, were considered. While care was taken to select headlines that dealt with a variety of topics, the majority of the headlines were related to COVID-19. This was due to the infodemic that ensued from the pandemic. The other topics included national policy, sustainability, crime and safety.

While the headlines were taken from Factually, they were adapted for the experiment. For example, we made grammar and wording modifications to achieve a consistent formal reporting style. We also changed some headlines to adjust their veracity for the experiment. For example, for News Item 9 in Table~\ref{tab:newsitems}, the original headline from Factually was “The Government has not proposed, planned nor targeted for Singapore to increase its population to 10 million” \citep{GovSG2020b}. We modified the headline to “The Government Has Proposed For Singapore To Increase Its Population To 10 Million”, keeping close to the original headline to ensure that no other changes in meaning were made. While we took news items from a threeyear period, we were aware that there might have been some developments in the respective event or phenomenon. As such, we checked each news at the time of the study to ensure that the facts applied even in 2021. For example, News Item 8 was fact checked on August 14, 2019 \citep{GovSG2019} and remained valid up to the time of the study.

\begin{table}[h]
    \centering
    \caption{The 16 pieces of news used in the experiment.}
    \label{tab:newsitems}
    \resizebox{\linewidth}{!}{
    \begin{tabular}{c p{0.95\linewidth} c}
        \toprule
        Item & Headline & \makecell{News\\veracity} \\
        \midrule
        1 & COVID-19 Does Not Spread Through Postal Articles & True \\
        2 & There Is Recourse In Law When There Has Been An Abuse Of POFMA Powers & True\\
        3  & Singapore Keeps Pace With Most Wealthy Developed Countries In Reducing Carbon Emissions Growth & True \\
        4 & Members of the Public Encouraged To Perform ``Hands-Only CPR'' Without The Need For Mouth-To-Mouth Breathing & True \\
        5 & Energy Transmission From TraceTogether Token Is Safe For Daily Use & True \\
        6 & MOE Spends More On Singaporean Students Than On Foreign Students & True \\
        7 & Safe Distancing Ambassadors Cannot Impose A Fine On Individuals Not Following Safe Distancing Laws & True \\
        8 & Voters Can Use A Taxi Or Private-Hire Vehicle To Travel To A Polling Station To Vote During The Election & True \\
        9 & The Government Has Proposed For Singapore To Increase Its Population To 10 Million & False \\
        10 & COVID-19 Vaccination Causes Stroke And Heart Attack & False \\
        11 & MOM States That All Employers Who Bring Their Foreign Workers For COVID-19 Testing Will Lose Their Work Pass Privileges & False \\
        12 & New Variant Of COVID-19 Found To Have Originated In Singapore & False \\
        13 & Police Officers Abuse Their Authority, Reprimanding And Taunting An Elderly Woman Who Did Not Have A Mask On & False \\
        14 & COVID-19 Tracker Has Been Secretly Installed On Every Phone And Can Be Found Under Phone Settings & False \\
        15 & The Government Has Spent Billions Of Taxpayers' Money On Extravagant Public Projects Like Jewel Changi Airport & False \\
        16 & People Are Robbing Residents Under The Pretext Of Distributing Masks, Purportedly Under A New Government Initiative & False \\
        \bottomrule
    \end{tabular}
    }
\end{table}

\section{Results}

In this section, we report the detailed statistical analysis conducted on the data. We then discuss the takeaways from the results in Sect.~\ref{discussion}.

\subsection{Statistical Analysis}

We analysed a total of 527 responses. In the counterbalancing of the Likert scale, 46.9 per cent of participants were assigned to the [False to True] scale and 53.1 per cent to the [True to False] scale. With the 16 news items as stimuli, the experiment data contained 527 × 16 = 8,432 trials. Repeated measures ANOVA with Greenhouse–Geisser corrections\footnote{Repeated measures ANOVA assumes sphericity by default, which is the condition that the variances of the differences between all possible pairs of a given within-subject independent variable are equal. If the assumption of sphericity is violated, we might end up with inflated $F$-scores and Greenhouse–Geisser corrections are applied to produce a more valid $F$-score.} where necessary were used to identify main effects for within-subject factors, followed by post hoc comparisons using pairwise t-tests with Bonferroni corrections\footnote{When performing post hoc analysis, we encounter the multiple comparisons problem. When using simultaneous statistical tests, each test has a potential to produce an effect, leading to Type I errors (incorrectly rejecting the null hypothesis, and thus incorrectly accepting an effect that is not there). To counter this issue, we use Bonferroni corrections, in which the $p$-value is multiplied by the number of pairwise comparisons to be made.\label{fn:bonf}} for interactions. We report the $F$-score, $p$-value and generalised eta squared value ($\eta^2_G$) for significant main effects, and the $p$-value for significant interactions. When describing the results, we report the mean ($M$) and median ($med$) as a measure of central tendency, and standard deviation ($SD$) and interquartile range ($IQR$) as a measure of spread.

\subsection{Accuracy of the Perceived Veracity}\label{acc}

The Accuracy of the Perceived Veracity (AccuracyPV) looks at how well participants performed in judging the veracity of news. It had a mean score of 2.68 ($med = 3, SD = 0.97, IQR = 1$).

\begin{figure}[h]
    \centering
    \begin{subfigure}[b]{0.49\linewidth}
        \centering
        \includegraphics[width=\linewidth]{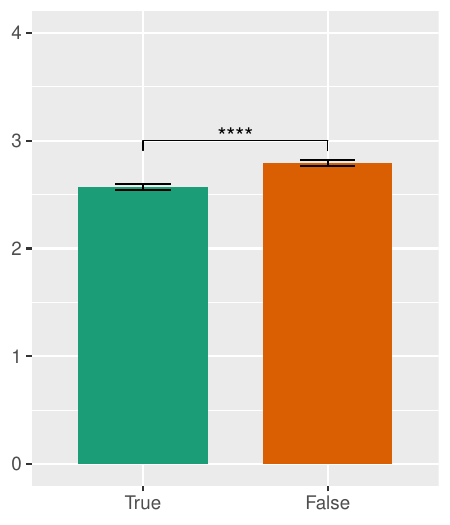}
        \caption{}
        \label{fig:accnv}
    \end{subfigure}
    \begin{subfigure}[b]{0.49\linewidth}
        \centering
        \includegraphics[width=\linewidth]{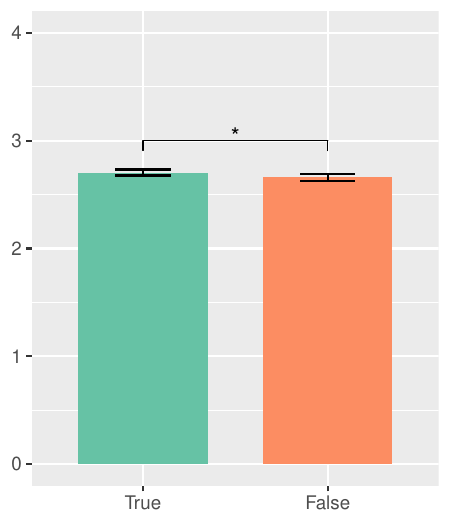}
        \caption{}
        \label{fig:accfcl}
    \end{subfigure}
    \begin{subfigure}[b]{0.49\linewidth}
        \centering
        \includegraphics[width=\linewidth]{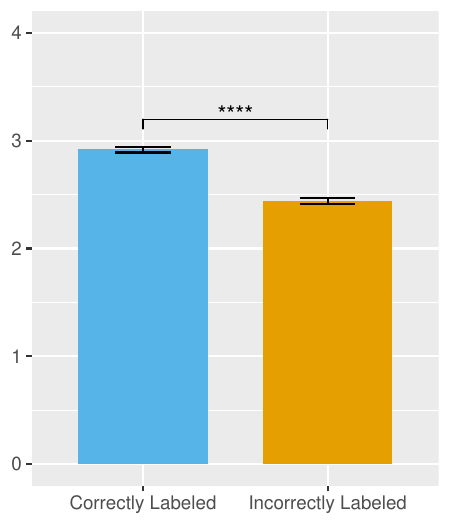}
        \caption{}
        \label{fig:acclp}
    \end{subfigure}
    \caption{Average AccuracyPV across different levels of (a) News Veracity, (b) Fact Check Label and (c) Label Precision. AccuracyPV scores are between 4 (high accuracy) and 1 (low accuracy). Error bars show 0.95 confidence intervals.}
    \label{fig:accuracy}
\end{figure}

\subsubsection{Main Effects}

The significant main effects of the independent variables are shown in Fig.~\ref{fig:accuracy}.

There was no significant main effect of Fact Checker on AccuracyPV ($p = 0.61$). The measured accuracy was highest for News Outlets ($M = 2.70, med = 3, SD = 0.95, IQR = 1$) and lowest for Control ($M = 2.67, med = 3, SD = 0.99, IQR = 1$).

There was a significant main effect of News Veracity on AccuracyPV ($F_{1,526}=34.15, p<0.0001, \eta^2_G=0.014$) where False news ($M = 2.79, med = 3, SD = 0.99, IQR = 2$) had higher accuracy than True news ($M = 2.57, med = 3, SD = 0.94, IQR = 1$) as shown in Fig.~\ref{fig:accnv}.

There was a significant main effect of Fact Check Label on AccuracyPV ($F_{1,526}=5.55, p=0.019, \eta^2_G=0.00052$) where news labelled True ($M = 2.70, med = 3, SD = 0.92, IQR = 1$) had higher accuracy than news labelled False ($M = 2.66, med = 3, SD = 1.01, IQR = 1$) as shown in Fig.~\ref{fig:accfcl}.

There was a significant main effect of Label Precision on AccuracyPV ($F_{1,526}=223.36, p<0.0001, \eta^2_G=0.062$) where Correctly Labelled news ($M = 2.92, med = 3, SD = 0.90, IQR = 2$) had higher accuracy than Incorrectly Labelled news ($M = 2.44, med = 2, SD = 0.97, IQR = 1$) as shown in Fig.~\ref{fig:acclp}.

\begin{figure}[h]
    \centering
    \begin{subfigure}[b]{0.85\linewidth}
        \centering
        \includegraphics[width=\linewidth]{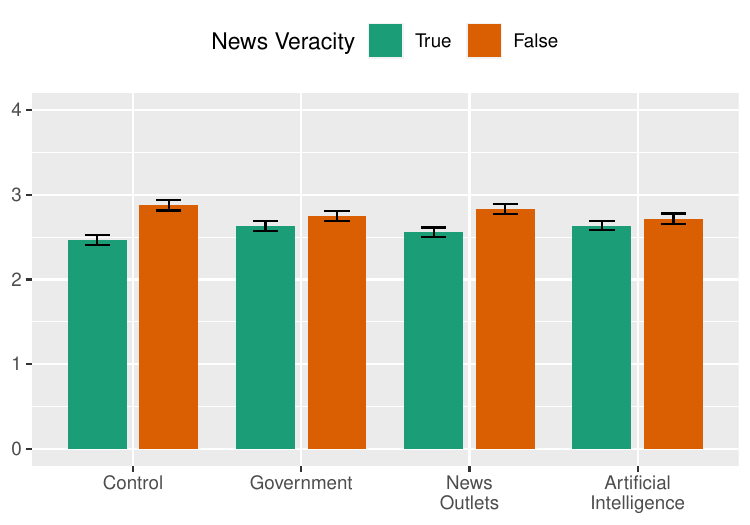}
        \caption{}
        \label{fig:accfcnv}
    \end{subfigure}
    \begin{subfigure}[b]{0.85\linewidth}
        \centering
        \includegraphics[width=\linewidth]{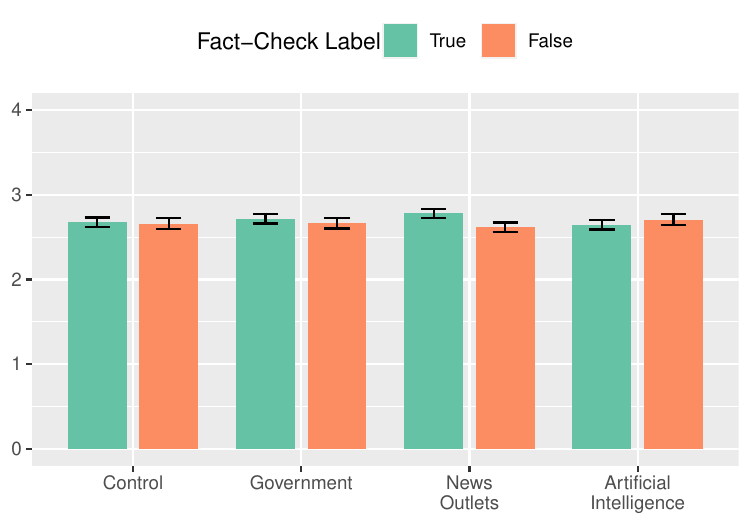}
        \caption{}
        \label{fig:accfcfcl}
    \end{subfigure}
    \vskip\baselineskip
    \begin{subfigure}[b]{0.85\linewidth}
        \centering
        \includegraphics[width=\linewidth]{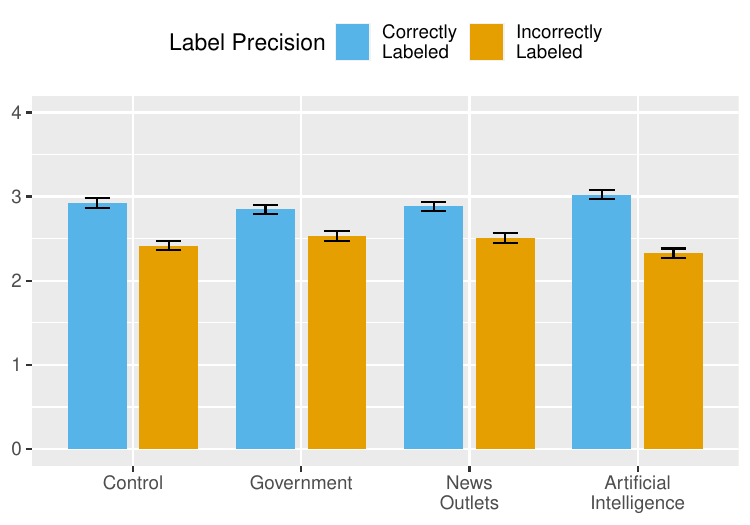}
        \caption{}
        \label{fig:accfclp}
    \end{subfigure}
    \caption{Average AccuracyPV across different levels of (a) Fact Checker and News Veracity, (b) Fact Checker and Fact Check Label and (c) Fact Checker and Label Precision. AccuracyPV scores are between 4 (high accuracy) and 1 (low accuracy). Error bars show 0.95 confidence intervals.}
    \label{fig:accuracyinteractions}
\end{figure}

\subsubsection{Interactions}

There was a Fact Checker × News Veracity interaction ($F_{3,1578}=21.03, p<0.0001, \eta^2_G=0.0050$). In Fig.~\ref{fig:accfcnv}, the gap was widest for the Control condition in which False news were identified more accurately by participants ($M = 2.88$) than True news ($M = 2.46$). The gap was overall smaller for the other conditions. Post hoc tests showed that Control × False achieved significantly higher accuracy than for Government and Artificial Intelligence (both $p < 0.001$). For True news, Control achieved a significantly lower accuracy compared to every other fact checker (all $p < 0.05$).

There was also a Fact Checker × Fact Check Label interaction ($F_{3,1578}=9.16, p<0.0001, \eta^2_G=0.0020$). Accuracy was slightly higher for news labelled as True rather than False for all the fact checkers except Artificial Intelligence (see Fig.~\ref{fig:accfcfcl}).

Lastly, there was a Fact Checker × Label Precision interaction ($F_{3,1578}=26.56, p<0.0001, \eta^2_G=0.0060$). While participants tended to be more accurate in judging the veracity of Correctly Labelled news than Incorrectly Labelled news (see Fig.~\ref{fig:accfclp}), the gap was widest for the Artificial Intelligence condition ($M = 3.02$ for Correctly Labelled and $M = 2.33$ for Incorrectly Labelled). From post hoc tests, accuracy was significantly higher for Artificial Intelligence × Correctly Labelled than every other fact checker (all $p < 0.01$), and significantly lower for Artificial Intelligence × Incorrectly Labelled than Government and News Outlets (both $p < 0.0001$).

\subsubsection{Addressing H1}

With no significant main effect of Fact Checker on the Accuracy of the Perceived Veracity, H1 (i.e., different fact checkers will lead to different levels of accuracy in judging the veracity of news) was not strongly supported. However, it was somewhat supported by the significant interactions of fact checker with the other independent variables. Considering News Veracity (see Fig.~\ref{fig:accfcnv}), the Control interface (that did not show a fact checker) is best avoided should the news be true as accuracy was significantly lower than for when the interface showed a fact checker. Considering the Label Precision (see Fig.~\ref{fig:accfclp}), Artificial Intelligence is most suitable for Correctly Labelled news as it had significantly higher accuracy than the other fact checkers, but should be avoided when news is Incorrectly Labelled, where it had lower accuracy than the other fact checkers instead.

\subsection{Adherence to the Fact Check Label}\label{adh}

The Adherence to the Fact Check Label (AdherenceFCL) measured how participants’ judgement of the veracity of news was affected by the fact check label. It had a mean score of 0.62 ($med = 1, SD = 0.49, IQR = 1$).

\begin{figure}[h]
    \centering
    \begin{subfigure}[b]{0.67\linewidth}
        \centering
        \includegraphics[width=\linewidth]{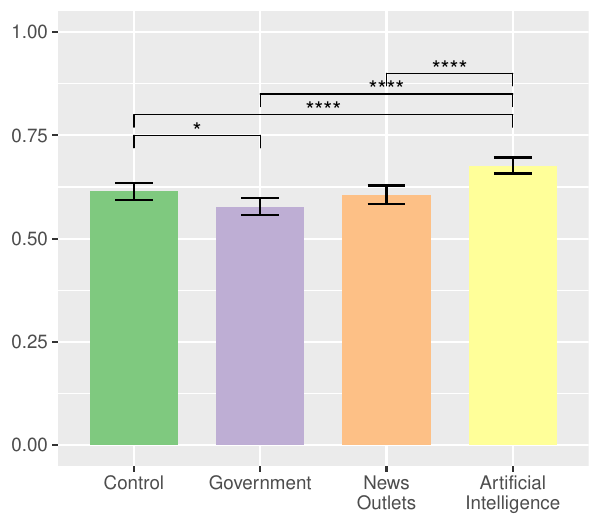}
        \caption{}
        \label{fig:adhfc}
    \end{subfigure}
    \begin{subfigure}[b]{0.49\linewidth}
        \centering
        \includegraphics[width=\linewidth]{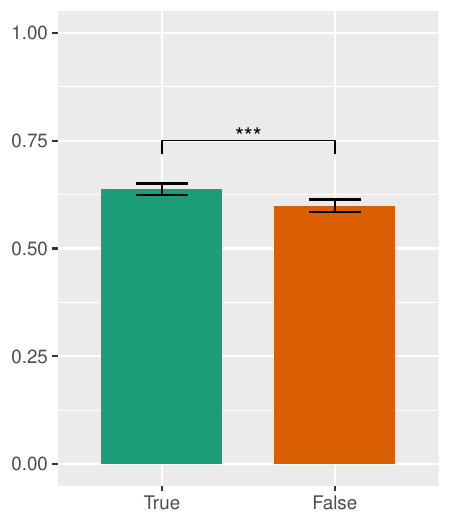}
        \caption{}
        \label{fig:adhnv}
    \end{subfigure}
    \begin{subfigure}[b]{0.49\linewidth}
        \centering
        \includegraphics[width=\linewidth]{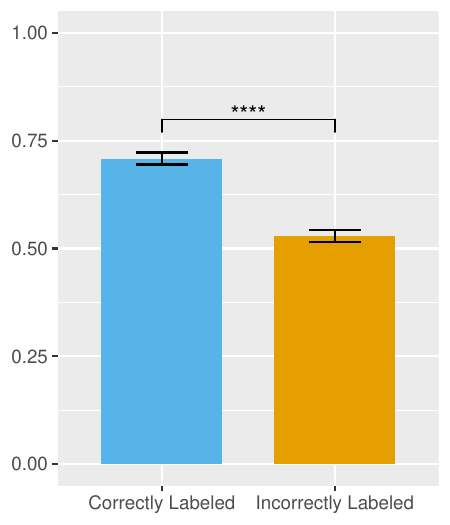}
        \caption{}
        \label{fig:adhlp}
    \end{subfigure}
    \caption{Average AdherenceFCL across different levels of (a) Fact Checker, (b) News Veracity and (c) Label Precision. AdherenceFCL scores are between 1 (high adherence) and 0 (low adherence). Error bars show 0.95 confidence intervals.}
    \label{fig:adherence}
\end{figure}

\subsubsection{Main Effects}

The significant main effects of the independent variables are shown in Fig.~\ref{fig:adherence}.

There was a significant main effect of Fact Checker on AdherenceFCL ($F_{3,1578}=25.39, p<0.0001, \eta^2_G=0.0058$) as shown in Fig.~\ref{fig:adhfc}. Significant pairwise comparisons were observed between Artificial Intelligence and the other fact checkers (all $p < 0.0001$), as well as between Government and Control ($p = 0.02$). Highest adherence was observed for Artificial Intelligence ($M = 0.68, med = 1, SD = 0.47, IQR = 1$), followed by Control ($M = 0.61, med = 1, SD = 0.49, IQR = 1$), News Outlets ($M = 0.61, med = 1, SD = 0.49, IQR = 1$), and Government ($M = .58, med = 1, SD = 0.49, IQR = 1$).

There was a significant main effect of News Veracity on AdherenceFCL ($F_{1,526}=16.77, p<0.0001, \eta^2_G=0.0017$) where True news ($M = 0.64, med = 1, SD = 0.48, IQR = 1$) had higher adherence than False news ($M = 0.60, med = 1, SD = 0.49, IQR = 1$) as shown in Fig.~\ref{fig:adhnv}.

There was no significant main effect of Fact Check Label on AdherenceFCL ($p = 0.29$). Similar levels of adherence were observed for news labelled True ($M = 0.61, med = 1, SD = 0.49, IQR = 1$) and False ($M = 0.63, med = 1, SD = 0.48, IQR = 1$).

There was a significant main effect of Label Precision on AdherenceFCL ($F_{1,526}=220.81, p<0.0001, \eta^2_G=0.035$) where Correctly Labelled news ($M = 0.71, med = 1, SD = 0.46, IQR = 1$) had higher adherence than Incorrectly Labelled news ($M = 0.53, med = 1, SD = 0.50, IQR = 1$) as shown in Fig.~\ref{fig:adhlp}.

\begin{figure}[h]
    \centering
    \begin{subfigure}[b]{0.85\linewidth}
        \centering
        \includegraphics[width=\linewidth]{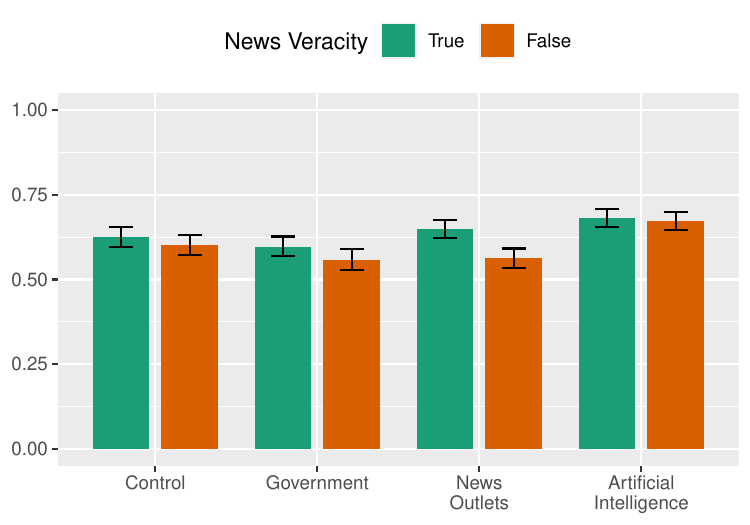}
        \caption{}
        \label{fig:adhfcnv}
    \end{subfigure}
    \begin{subfigure}[b]{0.85\linewidth}
        \centering
        \includegraphics[width=\linewidth]{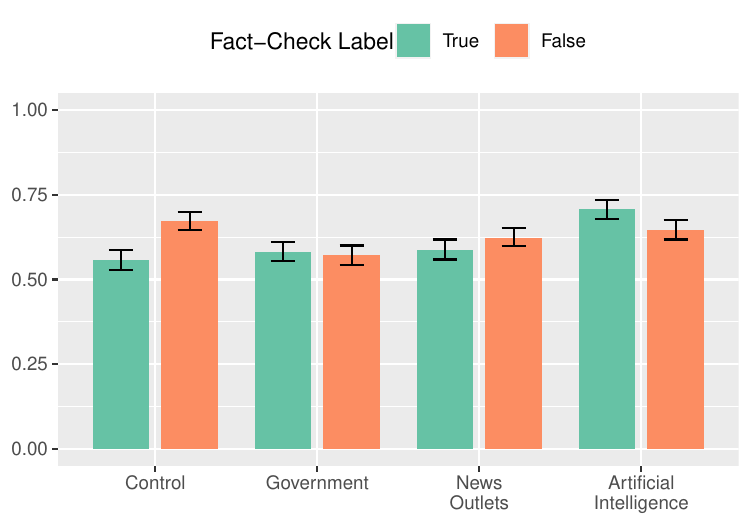}
        \caption{}
        \label{fig:adhfcfcl}
    \end{subfigure}
    \caption{Average AdherenceFCL across different levels of (a) Fact Checker and News Veracity, and (b) Fact Checker and Fact Check Label. AdherenceFCL scores are between 1 (high adherence) and 0 (low adherence). Error bars show 0.95 confidence intervals.}
    \label{fig:adherenceinteractions}
\end{figure}

\subsubsection{Interactions}

There was a significant Fact Checker × News Veracity interaction ($F_{3,1578}=3.67, p=0.012, \eta^2_G=0.00092$). From Fig.~\ref{fig:adhfcnv}, Artificial Intelligence had the highest adherence for both True news ($M = 0.68$) and False news ($M = 0.67$). For True news, there was significantly higher adherence for Artificial Intelligence than for every other fact checker (all $p < 0.0001$). For False news, there was significantly higher adherence for Artificial Intelligence than for Control and Government (both $p < 0.001$). The widest gap was observed in News Outlets ($M = 0.65$ for True news and $M = 0.56$ for False news).

There was also a significant Fact Checker × Fact Check Label interaction ($F_{3,1578}=17.85, p<0.0001, \eta^2_G=0.0046$). From Fig.~\ref{fig:adhfcfcl}, participants exhibited different behaviours in terms of adherence. They tended to follow Artificial Intelligence more for news labelled as True ($M = 0.71$) than for news labelled as False ($M = 0.48$) where there was significantly higher adherence for Artificial Intelligence × True than for every other fact checker (all $p < 0.0001$). Also, Control had the widest gap with adherence being higher for news labelled as False ($M = 0.47$) than for news labelled as True ($M = 0.56$). For False news, there was significantly higher adherence for Control than to Government and News Outlets (both $p < 0.01$).

\subsubsection{Addressing H2}

With a significant main effect of Fact Checker on the Adherence to the Fact Check Label (Fig.~\ref{fig:adhfc}), H2 (i.e., Different fact checkers will lead to different levels of adherence to the fact check labels when judging the veracity of news) is strongly supported. It comes as a surprise, however, that Government turned out to have the lowest adherence (i.e., degree of influence of the fact check label on participants’ judgement of the veracity of news). This runs contrary to our expectations, given people’s high trust in the Singapore government. Instead, Artificial Intelligence had the highest adherence and remained so in the interaction with News Veracity for both True and False news (see Fig.~\ref{fig:adhfcnv}). In the interaction with Fact Check Label, however, Artificial Intelligence was highest only for news labelled as True while Control was highest for news labelled as False (see Fig.~\ref{fig:adhfcfcl}). This suggests that Artificial Intelligence was more assuring than the other fact checkers for specifying True news while Control is more assuring for specifying False news.

\subsection{Performance in Perceived Veracity}

To understand the performance of participants in rating the veracity of news, we created a matrix of perceived veracity performance by the Adherence to the Fact Check Label. Table~\ref{tab:match} shows the percentage of responses to the authenticity question (out of 8,432 trials) in which the perceived veracity of news was right or wrong, based on whether the response adhered to the label. To obtain the right and wrong responses, we compared the perceived veracity response (from the authenticity question) with the actual veracity of the news (i.e., News Veracity). For instance, if the rating given by the participant was “True” or “Somewhat True” and the News Veracity of that news was “True”, there would be a match between the perceived veracity and the actual veracity of the news, indicating a right response. Conversely, if the News Veracity is “False”, there would not be a match, thereby indicating a wrong response.

\begin{table}[h]
    \centering
    \caption{Performance of participants’ perceived veracity of news by Adherence to the Fact Check Label in percentages ($N = 8,432$).}
    \label{tab:match}
    \begin{tabular}{cccc}
    \toprule
       &   & \multicolumn{2}{c}{\textbf{Adherence to the fact check label}} \\
       &   & Does adhere (\%) & Does not adhere (\%) \\
    \midrule
    \textbf{Perceived} & Right & 35.4 & 23.6 \\
    \textbf{veracity}   & Wrong & 26.4 & 14.6 \\
    \bottomrule
    \end{tabular}
\end{table}

\subsection{Post-Experiment Survey}\label{peq}

On a 7-point Likert scale (1: Strongly Disagree; 7: Strongly Agree), participants reported finding the chatbot interface easy to use ($M = 5.28, med = 5, SD = 1.23, IQR = 1.5$) and were open to having the chatbot available on their instant messaging apps ($M = 4.61, med = 5, SD = 1.54, IQR = 2$).

When asked to rank the fact checkers they preferred to provide fact checking services, the Singapore government was ranked first, followed by Singapore news outlets, and artificial intelligence systems. Significant differences for the rankings were found using the Friedman test\footnote{Friedman test is an alternative to repeated measures ANOVA which does not require normally distributed data, and is suitable for interval data, e.g., discrete scales like a Likert scale.} ($\chi^{2}(2)=467.43, p<0.0001, W=0.48$). Through post hoc analysis with Wilcoxon signed-rank tests with Bonferroni corrections\footnote{Wilcoxon signed-rank test is an alternative to paired t-tests, which does not require normally distributed data, and is suitable for interval data. The Bonferroni correction is applied to prevent the multiple comparisons problem as explained in footnote \ref{fn:bonf}.}, significant differences were found between all the fact checkers ($p < 0.0001$).

The same pattern was also reflected in how strongly participants reported to trust information from the fact checkers. They had strongest trust in the Singapore government ($M = 5.45, med = 6, SD = 1.36, IQR = 2$), followed by Singapore news outlets ($M = 5.01, med = 5, SD = 1.16, IQR = 2$), and artificial intelligence systems ($M = 4.51, med = 4, SD = 1.16, IQR = 1$).

\section{Discussion}\label{discussion}

In the results section, we discuss the research question and the implications of fact checkers and the fact checking chatbot as misinformation interventions on MIMS in general.

\subsection{Top Fact Checker}

To evaluate the effectiveness of the fact check labels, we assessed the Accuracy of the Perceived Veracity (see Sect.~\ref{acc}) and the Adherence to the Fact Check Label (see Sect.~\ref{adh}). Overall, the results veer towards artificial intelligence as the “top” fact checker, although we state this with reserve. While the fact checkers performed similarly in accuracy, they differed significantly in adherence. Artificial intelligence achieved highest adherence with most participants complying to its fact check labels (see Fig.~\ref{fig:adhfc}), suggesting that they found it more reliable. Taking a closer look at AccuracyPV, however, shows that while accuracy was highest for correctly labelled news, it was also lowest for incorrectly labelled news (see Fig.~\ref{fig:accfcnv}), implying that the use of artificial intelligence could be a double-edged sword. If the fact checker is truthful, artificial intelligence can enhance veracity perceptions of the information. Conversely, if the fact checker is mistaken or even deceptive, people may more easily fall for it instead. Nevertheless, this situation can be highly favourable by ensuring that the misinformation detection algorithm has outstanding performance and makes little or no errors. Artificial intelligence and machine learning solutions on misinformation detection have seen rapid development in recent years and are already used widely in social media platforms \citep{Twitter2018, Meta2020, Tiktok2020}. With more capable misinformation detection algorithms that are highly accurate, making them available on MIMS can bolster efforts in combating misinformation by addressing the scalability issue of fact checking \citep{Moy2021} where automated fact checking can address breaking news that professional human fact checkers have yet been able to review. In the current information scene where information is exchanged instantly and differs widely in context and content, being able to address misinformation with immediacy will be a valuable advantage.

\subsection{Contradiction Between Attitude and Behaviour}

From the fact checker ranking and reported trust results (see Sect.~\ref{peq}) where the government emerged as the top, followed by news outlets and artificial intelligence, the observation in the Adherence to the Fact Check Label by Fact Checker (see Sect.~\ref{adh}) where artificial intelligence had higher adherence ($M = 0.68$) than news outlets ($M = 0.61$) and the government ($M = 0.58$) is rather unexpected. Despite highest trust being reported in the government and lowest trust in artificial intelligence, fact check labels given by artificial intelligence were adhered to more strongly than fact check labels given by the government. This inconsistency suggests a contradiction between people’s attitude and behaviour towards fact checkers.

While the Singapore government and news outlets are seen as more trustworthy, this may be more the case when they are serving the role of a news provider rather than a news validator. The government has the standing and capacity to disseminate authoritative information, and news outlets have the responsibility to deliver information. Yet, Singapore news outlets have the reputation of being a fettered mouthpiece of the government among some segments of the population \citep{Aglionby2001, Hicks2013, TheIndependent2016}. Both government and news outlets fact checkers are thus tangled with perceptions of potential biases. The selected news items for the experiment had political undertones, and as such, government or perceived affiliated news outlets as fact checkers might have influenced people’s perceptions of the objectivity and truthfulness of the fact checks. The act of self fact checking (e.g., by the government) could have been perceived as less reliable than that by a third party, thereby diminishing trust and adherence towards the fact check labels provided by the government, despite the overall high trust in them. In contrast, artificial intelligence is computerised and may come off as being more objective and fairer \citep{Dijkstra1998}, resulting in its fact check labels being considered as more dependable instead \citep{Logg2019}. This is in line with the work of Araujo et al. which noted that “when respondents had to evaluate the potential fairness, usefulness and risk of specific decisions taken automatically by AI [Artificial Intelligence] in comparison to human experts, ADM [Automated Decision-Making] was often evaluated on par or even better for high-impact decisions” \citep{Araujo2020}.

\subsection{Efficacy vs. Blind Trust}

From Table~\ref{tab:match}, the collective performance of the participants in rating the veracity of news had only 59.0 per cent of responses being right and 41.0 per cent being wrong. Given that we deployed “opposite labels” such that half of the True news were labelled False, and half of the False news were labelled True, the poor performance might have been explained by participants basing their answers on the labels, particularly the incorrect ones. Indeed, this was the case with more than a quarter (26.4 per cent) of the perceived veracity responses being wrong as they adhered to the incorrect label. More broadly, 61.9 per cent of responses adhered to the label and 38.1 per cent did not. These observations suggest that participants depended on the fact checkers to provide accurate veracity ratings, perhaps when the news was novel or dubious to them. While the fact checking chatbot showed certain efficacy as the fact check labels were taken into account by the respondents, this also signals that people had some level of blind trust in the chatbot by treating the fact check labels as inherently accurate. This mirrors the observations of another study on automated fact checking \citep{Nguyen2018}. Thus, it is important for a fact checking service to uphold its integrity as it is relied upon by people to provide factual reporting that can be accepted without doubt.

\section{Limitations and Future Work}

\subsection{Measurement and Sample}

In the study, we used a 2-point scale for Adherence to the Fact Check Label as an indication of participants’ compliance to the fact check labels that could also have been indicative of a belief change from “False” to “True” and vice versa. However, it may be argued that a change in the degree of belief, such as from “Somewhat True” to “True” would also make for an effective fact check, and this could be captured using a more sensitive 4-point scale. While both are reasonable measures, we chose to use the 2-point scale as we were more interested in the polar switching of beliefs as we considered that one of the goals of a fact check is to convince people of the “truth” and align them with its absolute position.

While we sought to obtain a representative sample of the population for the study, our sample was skewed towards the young and middle age groups. Though we engaged a survey company for the recruitment of participants, the study was administered through an online web app in English. The elderly typically have lower digital literacy and may not be literate in English, and this posed some constraints in their recruitment. The elderly have been found to be more vulnerable to misinformation due to lacking technical skills and information literacy and their reliance on peers who may similarly lack expertise in identifying misinformation \citep{Yip2019, Ang2021}. While the younger population is not immune to misinformation, the older population is arguably the age group of greater concern. Future studies could translate the experiment in the vernacular languages to involve more elderly.

\subsection{Understanding the Contradiction}

One key finding of the study was the contradictory observation towards fact checkers where there was highest trust in yet lowest adherence to the government and the converse for artificial intelligence. While we sought to provide an explanation for the contradiction, this study did not explore the subtleties between the attitudes and behaviours of people towards fact checking and the fact checkers providing the service. More targeted investigations using both quantitative and qualitative methods are necessary to understand this observation.

Additionally, the political nature of a majority of the news headlines used in the study might have led participants to perceive the fact checks provided by the government as less trustworthy since they were self fact checks. This could have negatively affected the trustworthiness perceptions of government fact checks despite the overall high trust in the government and thus resulted in lower adherence. For investigators interested in conducting similar work in the future that may involve stakeholders with perceived vested interests, we advise using news from a variety of political and non-political topics (e.g., sports, entertainment and science).

\subsection{Beyond Textual Misinformation and Fact Checks}

Misinformation can take on many forms on MIMS. The various multimedia functionalities have given rise to a variety of formats for information to be exchanged such as through text, image, video and audio content. There are also more complicated text messages such as chain letter style messages containing partially or entirely fabricated content \citep{Lomas2018}. Fake images and videos are also of concern as visuals can be more convincing \citep{Swayne2021}. While audio-based misinformation is rarer, the voice messaging function can foster its spread since it is an easy-to-use function that may appeal more to the older or less tech-savvy people who are also more vulnerable. Misinformation in the audio form on MIMS, however, remains largely unstudied. In a similar vein, many fact checks are text-based. Alternative mediums like image, video and audio may deliver more entertaining, convincing and effective fact checking \citep{Pasquetto2020} that could generate greater interest and reach than the misinformation. Using multimedia may also prevent the chatbot from becoming dull and sustain users’ interest.

\section{Conclusion}

Singapore is a culturally diverse and digitally connected country where citizens have high confidence in the government. As misinformation permeates in MIMS that are used widely for personal communication, misinformation has become a greater threat, particularly to the older population who as non-digital natives are inevitably naiver to the pitfalls of the chaotic information landscape and are more susceptible. In seeking to understand measures that directly mitigate misinformation, our experimental study investigated the effectiveness of a fact checking chatbot misinformation intervention and the effect that trust in fact checkers providing the service have on the perceived veracity of news. A major finding of the study was the contradiction observed between participants’ trust in the fact checkers and the reliance on them when rating the veracity of news. News and consequently misinformation relating to government activities that are of public interest often emerge and yet, fact checks from the government are more likely to be dismissed compared to that of other fact checkers. This brings us to question the practicality of government fact checkers, and in a broader sense, of self fact checkers. On one hand, transparency about the fact checking process could help ameliorate concerns regarding the fact checker \citep{Humprecht2020}, yet on the other, resources could be better diverted to third party fact checkers. Our study points to the potential of artificial intelligence fact checkers instead. In the future, extending this work to other multimedia forms of misinformation and fact checks will contribute to the development of the chatbot intervention in terms of usability and sustainability.

\bibliographystyle{ACM-Reference-Format}
\bibliography{sample-base}

\end{document}